\documentclass[english]{article}
\usepackage[T1]{fontenc}
\usepackage[latin9]{inputenc}
\usepackage{amsmath}
\usepackage{amssymb}
\usepackage{stackrel}
\usepackage{graphicx}

\makeatletter
\newcommand{\lyxaddress}[1]{
\par {\raggedright #1
\vspace{1.4em}
\noindent\par}
}

\@ifundefined{date}{}{\date{}}
\makeatother

\usepackage{babel}
\begin{document}

\title{Geodesics around oscillatons made of exponential scalar field potential }

\author{\textit{Ali. Mahmoodzadeh}\thanks{a.mahmoodzadeh130@gmail.com}\textit{
, B. Malakolkalami}\thanks{B.Malakolkalami@uok.ac.ir}}
\maketitle

\lyxaddress{Faculty of Science, University of Kurdistan, Sanandaj, P. O. Box
:416, Iran, Tel : +988716660066}
\begin{abstract}
Oscillatons are spherically symmetric solutions to the Einstein-Klein-Gordon
(EKG) equations for soliton stars made of real time-dependent scalar
fields. These equations are non singular and satisfy flatness conditions
asymptotically with periodic time dependency. In this paper, we investigate
the geodesic motion of particles moving around an oscillaton related
to a time-dependent scalar field. Bound orbital is found for these
particles under the condition of particular values of angular momentum
$L$ and initial radial position $r_{0}$. We discuss this topic for
an exponential scalar field potential which could be of the form of
$V(\Phi)=V_{0}e^{\lambda\sqrt{k_{0}}\Phi}$ with a scalar field $\Phi(r,t)$
and investigate whether the radial coordinates of such particles oscillate
in time or not and thereby we could predict the corresponding oscillating
period as well as amplitude. It is necessary to recall, in general
relativity, a geodesic generalizes the notion of a \textquotedbl{}straight
line\textquotedbl{} to curved space-time. Importantly, the world line
of a particle free from all external, non-gravitational forces, is
a particular type of geodesic. In other words, a freely moving or
falling particle always moves along a geodesic. In general relativity,
gravity can be regarded as not a force but a consequence of a curved
space-time geometry where the source of curvature is the stress\textendash energy
tensor (representing matter, for instance). Thus, for example, the
path of a planet orbiting around a star is the projection of a geodesic
of the curved 4-D space-time geometry around the star onto 3-D space.

\vspace{1cm}
\end{abstract}
\textbf{keywords  }Einstein-Klein-Gordon equations, metric functions,
geodesics \vspace{2cm}

\textbf{\large{}1. Introduction} 

As we know the most accepted cosmological models for understanding
the mechanism of structure formation are those that contain dark matter
and dark energy, with a possibly negative equation of state as the
main constituents of the Universe. Both dark matter and dark energy
can be described by a \textit{dynamical scalar field $\Phi$} that
rolls in its potential $V(\Phi)$ {[}1,2{]}. On the other hand, nowadays
oscillatons are well known as astronomical objects with a spherically
symmetric and time-dependent metrics which are asymptotically flat
and non singular solutions to the Einstein-Klein-Gordon (EKG) equations
and their properties have been the subject of recent studies as a
possibility for the role of dark matter {[}3-6{]}. From astrophysical
point of view oscillatons are gravitationally bound objects with the
real scalar field $\Phi$ endowed with a scalar field potential $V(\Phi)$
which can be the subject of scalar field dark matter(SFDM) hypothesis
at the \textit{galactic scale} {[}7-9{]}. First of all we know that
the space-time around an oscillaton varies with time depending on
the form of the scalar field and scalar field potential. But on the
other side we know that there is not yet a general agreement on the
correct form that this scalar potential $V(\Phi)$ should have, see
for instance {[}1{]} and references there in. In some previous works,
for instance{[}10{]}, has been suggested an exponential\textendash like
scalar field potential which fits very well the constraints from big
bang nucleosynthesis, due to its (so called) tracker solutions, and
then fine tuning can be avoided. The most simple example of a potential
with both exponential behavior and a minimum is a \textit{$cosh$}
potential{[}11{]}. Recently we propose another form of the scalar
field potential, described by $V(\Phi)=V_{0}e^{\lambda\sqrt{k_{0}}\Phi}$,{[}12{]}
where $V_{0}$ and $\lambda$ are free parameters which should be
fixed through the cosmological observation {[}13{]} or constraints
which are imposed through the formulation of the problem{[}12{]}.
Thus the geodesic motion of real particles around an oscillaton varies
depending on the form of the potential. In previous works, the geodesics
around an oscillaton made out of a quadratic $V(\Phi)=\frac{1}{2}m_{\varPhi}^{2}\Phi^{2}$
scalar field potential has been carried out{[}14{]} and in this paper
we are interested in to do this for the new exponential form of the
potential mentioned above. Second if we assume that a galaxy is basically
made out of scalar field dark matter described by an oscillaton, one
would hope to observe periodic variations of stars$^{,}$ path around
the galaxy. These two topics are the main motivations for studying
what we are interested in. A summary of this study is as follows:

In section two we review the mathematical background of oscillatons
endowed with an exponential scalar field potential $V(\Phi$). In
third section we solve the geodesic equations numerically and investigate
their properties. In the last part we summarize the results and make
some final comments for next investigations.

\vspace{1cm}

\textbf{\large{}2. Mathematical background of oscillatons}\textbf{\vspace{0.5cm}
}

Oscillatons are time dependent spherically symmetric and asymptotically
flat solutions for the coupled Einstein-Klein-Gordon (EKG) equations.
Numerical solutions have been found for these equations by using the
Fourier expansions. But non-linearity of these equations has caused
only few modes have been employed {[}13, 16{]}. The most general spherically
symmetric metric is written as 

\begin{flushleft}
\hspace{20bp}$ds^{2}=g_{\alpha\beta}dx^{\alpha}dx^{\beta}=-e^{\nu-\mu}dt^{2}+e^{\nu+\mu}dr^{2}+r^{2}(d\theta^{2}+sin^{2}\theta d\varphi^{2})$
, \hspace{32bp}(1)
\par\end{flushleft}

where $\nu=\nu(t,r)$ and $\mu=\mu(t,r)$ are functions of time and
radial position (the units are so chosen in which$\hbar=c=1$). The
energy-momentum tensor endowed for a real scalar field $\Phi(t,r)$
with a scalar field potential V($\Phi$) is defined as {[}9{]}

\begin{flushleft}
\hspace{70bp}$T$$_{\alpha\beta}=\Phi_{,\alpha}\Phi_{,\beta}-\frac{1}{2}g_{\alpha\beta}${[}$\Phi^{,\gamma}\Phi_{,\gamma}$+$2V$($\Phi)]$.\hspace{92bp}(2)
\par\end{flushleft}

The non-vanishing components of $T_{\alpha\beta}$ are

\begin{flushleft}
\hspace{63bp}$-T^{0}$$_{0}=\rho_{\Phi}=\frac{1}{2}[e^{-(\nu-\mu)}\overset{.}{\Phi^{2}}+$$e$$^{-(\nu+\mu)}\Phi^{'2}+2V(\Phi)]$,\hspace{40bp}(3.
a)
\par\end{flushleft}

\begin{flushleft}
\hspace{70bp}$T_{01}=p_{\Phi}=\overset{.}{\Phi}\Phi^{'}$,\hspace{172bp}
(3. b)
\par\end{flushleft}

\begin{flushleft}
\hspace{70bp}$T^{1}$$_{1}=p_{r}=\frac{1}{2}${[}$e^{-(\nu-\mu)}\overset{.}{\Phi^{2}}+$$e$$^{-(\nu+\mu)}\Phi^{'2}-2V(\Phi)]$,\hspace{42bp}(3.
c)
\par\end{flushleft}

\begin{flushleft}
\hspace{70bp}$T^{2}$$_{2}=p_{\bot}$=$\frac{1}{2}${[}$e^{-(\nu-\mu)}\overset{.}{\Phi^{2}}-$$e$$^{-(\nu+\mu)}\Phi^{'2}-2V(\Phi)]$,\hspace{45bp}(3.
d)
\par\end{flushleft}

and we have also $T^{3}$$_{3}=T^{2}$$_{2}$. Over dots denote $\frac{\partial}{\partial t}$
and primes denote $\frac{\partial}{\partial r}$ . The different components
of the tensor for the scalar field mentioned above are identified
as the energy density, the momentum density, the radial pressure and
the angular pressure respectively. Einstein equations, $G$$_{\alpha\beta}=R_{\alpha\beta}-\frac{1}{2}g_{\alpha\beta}R$=$k_{0}T_{\alpha\beta}$
are used to obtain differential equations for functions $\nu$, $\mu$
, then 

\begin{flushleft}
\hspace{110bp}$(\nu+\mu)^{.}=k_{0}r\overset{.}{\Phi}\Phi^{'}$,\hspace{126bp}(4.
a)
\par\end{flushleft}

\begin{flushleft}
\hspace{110bp}$\nu^{'}=\frac{k_{0}r}{2}$($e$$^{2\mu}\overset{.}{\Phi^{2}}$+$\Phi^{'^{2}}$)
,\hspace{111bp}(4. b)
\par\end{flushleft}

\begin{flushleft}
$\hspace{110bp}\mu^{'}=\frac{1}{r}[1+$$e$$^{\nu+\mu}($$k$$_{0}r$$^{2}V$($\Phi)$-$1$$)]$
,\hspace{77bp}(4. c)
\par\end{flushleft}

where $R_{\alpha\beta}$ and $R$ are the Ricci tensor and Ricci scalar
respectively and $k_{0}=8\pi G=\frac{8\pi}{m_{pl}^{2}}$ ,where the
gravitational constant, $G$ , is the inverse of the reduced Planck
mass, $m_{pl}$,  squared. The conservation of the scalar field tensor
reads as 

\begin{flushleft}
\hspace{110bp}$T_{;\beta}^{\alpha\beta}=\Phi^{,\alpha}[\square\Phi-\frac{dV(\Phi)}{d\Phi}]=0$
,\hspace{93bp}(5)
\par\end{flushleft}

where $\square=\partial_{\alpha}\partial^{\alpha}=g_{\alpha\beta}\partial^{\alpha}\partial^{\beta}(\alpha,$
$\beta$=$0$, $1$ ,$2$, $3$) is the d$^{,}$Alembertian operator.
Therefore we can obtain the Klein-Gordon (KG) equation for the scalar
field $\Phi(t,r)$,

\begin{flushleft}
\hspace{80bp}$\Phi^{''}+\Phi^{'}(\frac{2}{r}-\mu^{'})-e^{\nu+\mu}\frac{dV(\Phi)}{d\Phi}=e^{2\mu}(\overset{..}{\Phi}+\overset{.}{\mu}\overset{.}{\Phi})$.\hspace{56bp}(6)
\par\end{flushleft}

If we choose $\Phi(r,t)=\sigma(r)\phi(t)$ , then Eq. (6) reads as

\begin{flushleft}
\hspace{20bp}$\phi\{\sigma^{''}+\sigma^{'}(\frac{2}{r}-\mu^{'})\}-\lambda\sqrt{k_{0}}V_{0}e^{\nu_{0}+\mu_{0}}e^{\lambda\sqrt{k_{0}}\sigma(r)\phi(t)}=e^{2\mu}\sigma(\overset{..}{\phi}+\overset{.}{\mu}\overset{.}{\phi})$.\hspace{28bp}(7)
\par\end{flushleft}

The following Fourier expansions 

\begin{flushleft}
\hspace{50bp}$e^{\pm f(x)cos(2\theta)}=I_{0}(f(x))+2\stackrel[n=1]{\infty}{\sum}(\pm1)^{n}I_{n}(f(x))cos(2n\theta)$
,\hspace{37bp}(8)
\par\end{flushleft}

where $I_{n}(z)$ are the modified Bessel functions of the first kind,
require that Eq. (7) changes into

\begin{flushleft}
$\frac{1}{\sigma}\{\sigma^{''}+\sigma^{'}(\frac{2}{r}-\mu^{'})-\lambda\sqrt{k_{0}}V_{0}e^{\nu+\mu}(I_{0}(\lambda\sqrt{k_{0}}\sigma)+2I_{1}(\lambda\sqrt{k_{0}}\sigma))\}=\frac{e^{2\mu}}{\phi}(\overset{..}{\phi}+\overset{.}{\mu}\overset{.}{\phi})+\frac{\lambda\sqrt{k_{0}}V_{0}e^{\nu+\mu}}{\sigma\phi}[I_{0}(\lambda\sqrt{k_{0}}\sigma)+2I_{1}(\lambda\sqrt{k_{0}}\sigma)]$
\hspace{103bp}(9)
\par\end{flushleft}

It is necessary to recall that , we have taken into account only the
first term of the exponential field expansion for simplicity to obtain
Eq. (9). If we choose

\begin{flushleft}
\hspace{110bp}$\sqrt{k_{0}}\Phi(r,t)=2\sigma(r)cos(\omega t)$ ,\hspace{96bp}(10)
\par\end{flushleft}

where $\omega$ is the fundamental frequency of the scalar field,
then integrating on the Eq. (4.a) formally yields

\begin{flushleft}
\hspace{100bp}$\nu+\mu=(\nu+\mu)_{0}+r\sigma\sigma^{'}cos(2\omega t)$,\hspace{81bp}(11)
\par\end{flushleft}

where $(\nu+\mu)_{0}$ is a function of r-coordinate only. The boundary
conditions which will be discussed latter require that

\begin{flushleft}
\hspace{130bp}$\nu_{1}+\mu_{1}=r\sigma\sigma^{'}$.\hspace{128bp}(12)
\par\end{flushleft}

It is useful to perform the following variable changes for numerical
purposes of the following form

\begin{flushleft}
\hspace{40.5bp}$x=m_{\Phi}r$ ,\hspace{15bp}$\varOmega=\frac{\omega}{m_{\Phi}}$
,\hspace{15bp}$e^{\nu_{0}}\rightarrow$$\varOmega e^{\nu_{0}}$,\hspace{15bp}$e^{\mu_{0}}\rightarrow\varOmega^{-1}e^{\mu_{0}}$.\hspace{38bp}(13) 
\par\end{flushleft}

Hence by using Eqs. (4.a-6), (10, 12) and setting each Fourier component
to zero the metric functions and the radial part of the scalar field
are obtained as

\begin{flushleft}
$\nu_{0}^{'}=x[e^{2\mu_{0}}\sigma^{2}\left(I_{0}(2\mu_{1})-I_{1}(2\mu_{1})\right)+\sigma^{'2}]$
,\hspace{146bp}(14)
\par\end{flushleft}

\medskip{}

\begin{flushleft}
$\nu_{1}^{'}=x[e^{2\mu_{0}}\sigma^{2}\left(2I_{1}(2\mu_{1})-I_{0}(2\mu_{1})-I_{2}(2\mu_{1})\right)+\sigma^{'2}]$
,\hspace{97bp}(15)
\par\end{flushleft}

\medskip{}

\begin{flushleft}
$\mu_{0}^{'}=\frac{1}{x}\{1+e^{\nu_{0}+\mu_{0}}[x^{2}\left(I_{0}(x\sigma\sigma^{'})I_{0}(2\lambda\sigma)+2I_{1}(x\sigma\sigma^{'})I_{2}(2\lambda\sigma)\right)-I_{0}(x\sigma\sigma^{'})]\}$,(16)
\par\end{flushleft}

\medskip{}

\begin{flushleft}
$\mu_{1}^{'}=\frac{2}{x}e^{\nu_{0}+\mu_{0}}[x^{2}\left(I_{0}(x\sigma\sigma^{'})I_{2}(2\lambda\sigma)+I_{1}(x\sigma\sigma^{'})I_{0}(2\lambda\sigma)+I_{2}(X\sigma\sigma^{'})I(2\lambda\sigma)\right)-I_{1}(x\sigma\sigma^{'})]$
,\hspace{279bp}(17)
\par\end{flushleft}

\medskip{}

\begin{flushleft}
$\sigma^{''}=-\sigma^{'}(\frac{2}{x}-\mu_{0}^{'}-\frac{1}{2}\mu_{1}^{'})+\lambda e^{\nu_{0}+\mu_{0}}I_{0}(2\lambda\sigma)\left(I_{0}(x\sigma\sigma^{'})+I_{1}(x\sigma\sigma^{'})\right)-e^{2\mu_{0}}\sigma[I_{0}(2\mu_{1})(1-\mu_{1})+I_{1}(2\mu_{1})+I_{2}(2\mu_{1})]$
.\hspace{138bp}(18)
\par\end{flushleft}

\medskip{}

Now primes denote $\frac{d}{dx}$. The two last equations are true
if we put , $k_{0}V_{0}=m_{\Phi}^{2}$. With initial boundary values
for $\nu_{0}(x=0)$ , $\nu_{1}(x=0)$ , $\mu_{0}(x=0)$ , $\mu_{1}(x=0)$
, $\sigma(x=0)$ and parameter $\lambda$, these equations are soluble
numerically and the answers are in the form of Interpolation Functions.
Therefore the metric functions and then the metric coefficients are
completely determined. Singularity points for the metric functions
are also obtained through boundary conditions. For example if we put
$\nu_{0}(x=0)=-0.11$ , $\nu_{1}(x=0)=0.17$ , $\mu_{0}(x=0)=0.11$
, $\mu_{1}(x=0)=-0.17$ , $\sigma(x=0)=0.3$ , and $\lambda=-0.1$
then singularity happens at point $x=1.62380$. ( with increasing
the values of $\lambda$ and $\sigma$, singularity points will happen
at less and less values of $x$ ). The important point that should
be mentioned here is that the exponential and quadratic scalar potentials
described by $V(\Phi)=V_{0}e^{\lambda\sqrt{k_{0}}\varPhi}$ and $V(\varPhi)=\frac{1}{2}m_{\varPhi}^{2}\varPhi^{2}$
respectively with the same initial boundary conditions cause different
metric functions and metric coefficients{[}4,12,16{]}. 

For next purposes the answers obtained from Eqs. (14-17) are fitted
to six order which are in compliance with equations (19-22). The results
of these fittings are shown in Fig. 1. From now on the metric functions
are presented by these fitted functions of the following form:

\bigskip{}

$\nu_{0}(x)=-0.106045-0.35163x+3.1483x^{2}-9.5576x^{3}+13.2842x^{4}-8.45323x^{5}+2.01531x^{6}$,\hspace{280bp}(19)

\bigskip{}

$\nu_{1}(x)=0.171322-0.12740x+1.08631x^{2}-3.76961x^{3}+5.48473x^{4}-3.68322x^{5}+0.93388x^{6}$,\hspace{280bp}(20)

\bigskip{}

$\mu_{0}(x)=0.178029-1.72671x+12.168x^{2}-33.2923x^{3}+44.0517x^{4}-27.2995x^{5}+6.44389x^{6}$,\hspace{280bp}(21)

\bigskip{}

$\mu_{1}(x)=0.169913-0.008487x+0.096403x^{2}-0.242323x^{3}+0.32989x^{4}-0.202957x^{5}+0.0397515x^{6}.$\hspace{211bp}(22)

\vspace{0.5cm}

\includegraphics[width=6cm,height=5cm]{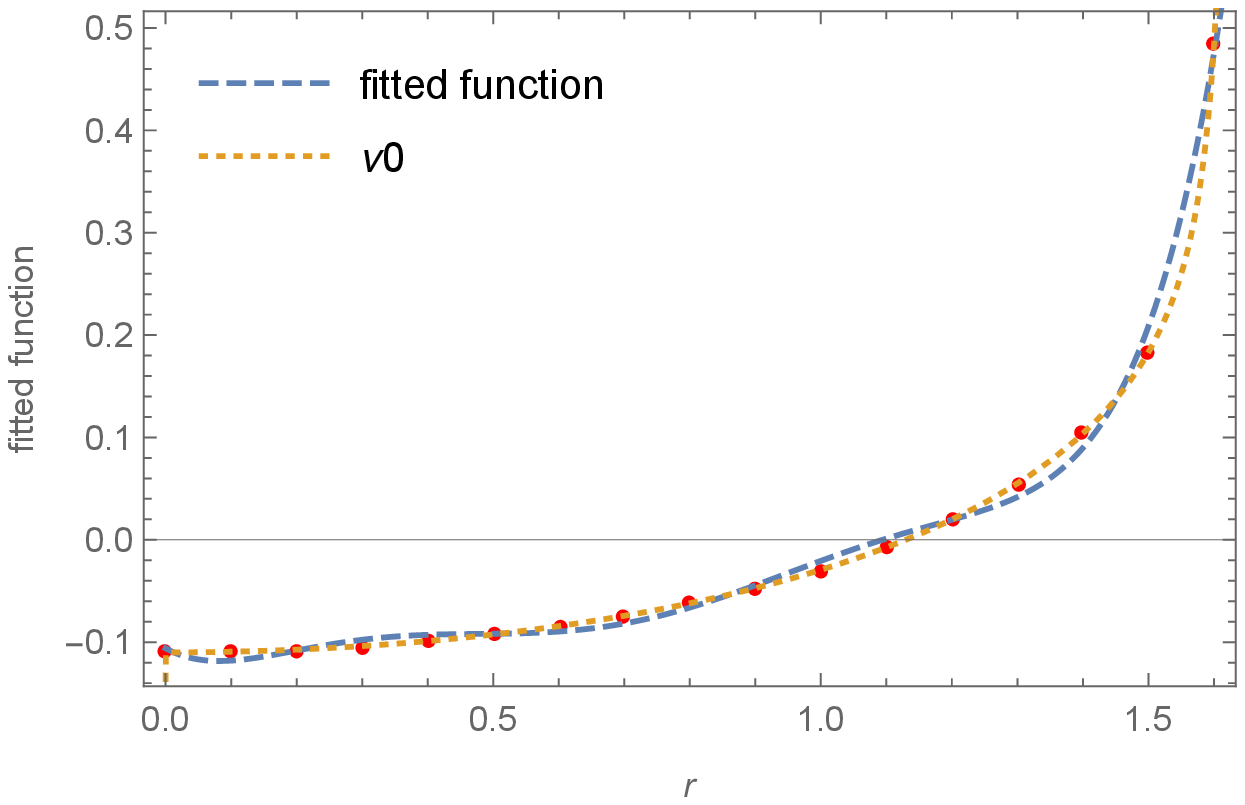}\includegraphics[width=6cm,height=5cm]{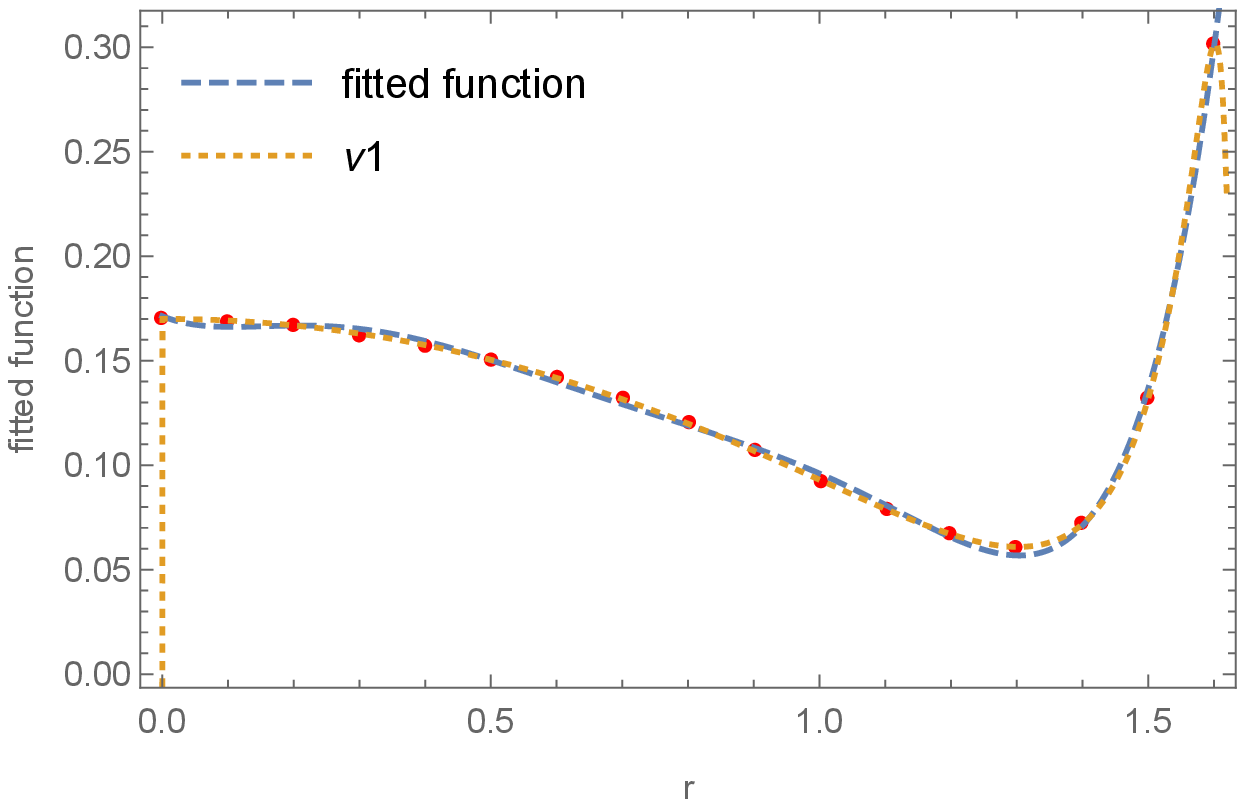}

\vspace{0.5cm}

\includegraphics[width=6cm,height=5cm]{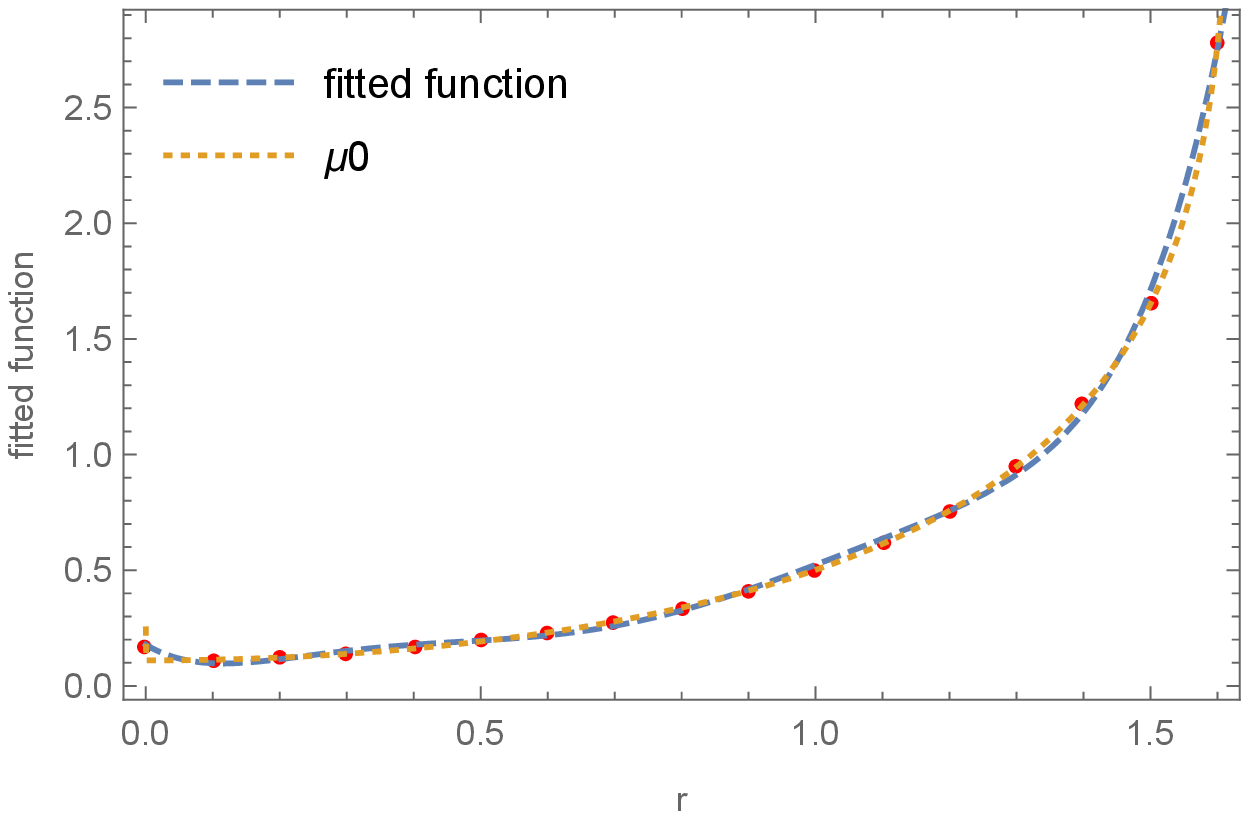}\includegraphics[width=6cm,height=5cm]{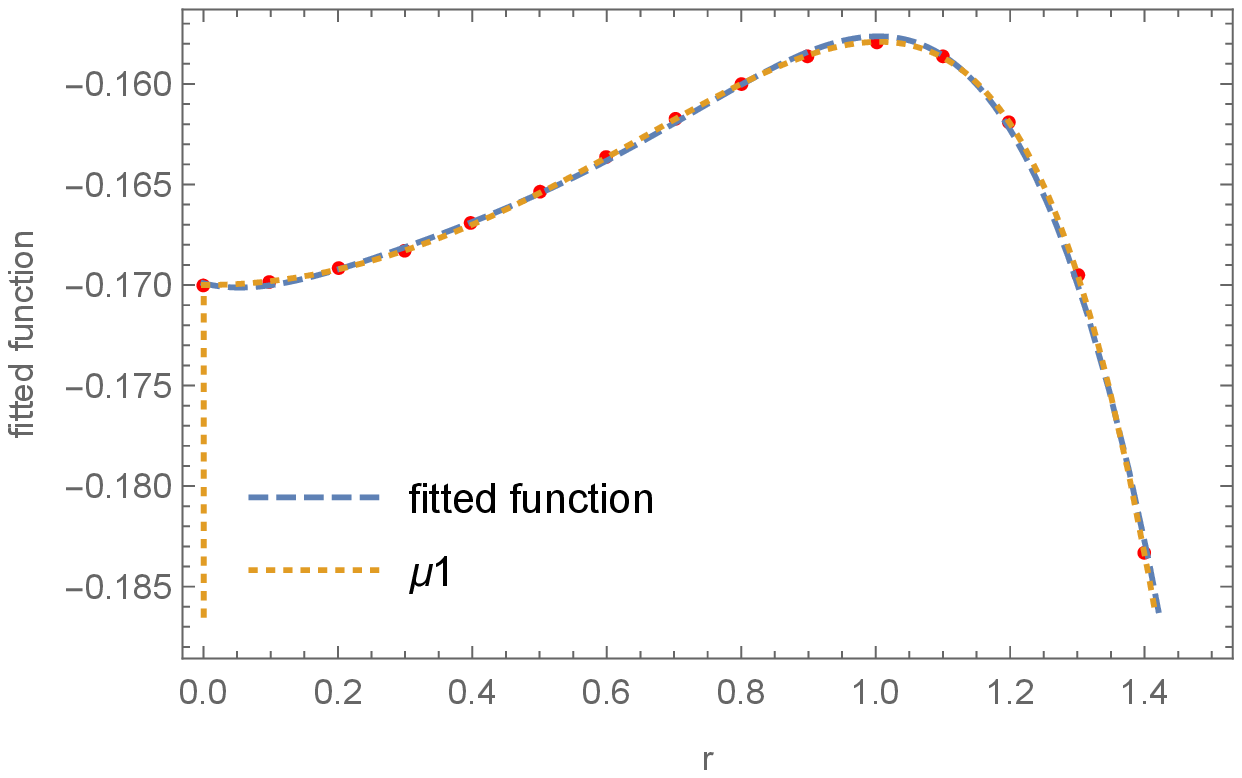}

Fig. 1. The metric functions and their corresponding fitted functions.

\vspace{1cm}

The next step is to specify the geodesics around oscillatons for real
particles. As we mentioned above, this work for a real particles around
oscillatons described by a quadratic scalar field potential has been
done in previous work{[}14{]} and investigation of geodesics for real
particles around oscillatons made of an exponential scalar field potential
discribed by{[}12{]}, is the subject of this paper. For this work
we continue the process as follow:

If we redefine the metric coefficients as the following form

\begin{flushleft}
\hspace{10bp}$g_{tt}=-e^{\nu-\mu}=-B(x,t)$ ,\hspace{14bp}$g_{rr}=e^{\nu+\mu}=A(x,t)$,\hspace{14bp}$C(x,t)=\frac{A(x,t)}{B(x,t)}$,\hspace{16bp}(23)
\par\end{flushleft}

then the Einstein equations can be rewritten as 

\begin{flushleft}
\hspace{60bp}$\frac{\partial A}{\partial t}=k_{0}A\frac{\partial\Phi}{\partial t}\frac{\partial\Phi}{\partial x}$
,\hspace{178bp}(24. a)
\par\end{flushleft}

\begin{flushleft}
\hspace{60bp}$\frac{\partial A}{\partial x}=\frac{k_{0}Ax}{2}[C(\frac{\partial\Phi}{\partial t})^{2}+(\frac{\partial\Phi}{\partial x})^{2}+2AV$($\Phi)]$
,\hspace{80bp}(24. b)
\par\end{flushleft}

\begin{flushleft}
\hspace{60bp}$\frac{\partial C}{\partial x}=\frac{2C}{x}[1+A(k_{0}x^{2}V(\Phi)-1]$
,\hspace{118bp}(24. c)
\par\end{flushleft}

\begin{flushleft}
\hspace{60bp}$C\frac{\partial^{2}\Phi}{\partial t^{2}}=-\frac{1}{2}\frac{\partial C}{\partial t}\frac{\partial\Phi}{\partial t}+\frac{\partial^{2}\Phi}{\partial x^{2}}+\frac{\partial\Phi}{\partial x}[\frac{2}{x}-\frac{1}{2C}\frac{\partial C}{\partial x}]-A\frac{dV(\Phi)}{d\Phi}$
.\hspace{23bp}(24. d)
\par\end{flushleft}

It is clear that by combining of Eq. (24. c) and (24. d) the non-linearity
of these equations can be minimized. By applying $A(x,t)$ , $B(x,t)$
and $C(x,t)$ the non-vanishing components of $T_{\alpha\beta}$ can
be renovated as 

\begin{flushleft}
$\hspace{60bp}-T_{0}^{0}=\rho_{\Phi}=\frac{1}{2}[\frac{C(x,t)}{A(x,t)}(\frac{\partial\Phi}{\partial t})^{2}+\frac{1}{A(x,t)}(\frac{\partial\Phi}{\partial x})^{2}+2V(\Phi)]$
,\hspace{30bp}(25. a)
\par\end{flushleft}

\begin{flushleft}
\hspace{60bp}$T_{1}^{0}=p_{\Phi}=(\frac{\partial\Phi}{\partial t})(\frac{\partial\Phi}{\partial x})$
,\hspace{156bp}(25. b)
\par\end{flushleft}

\begin{flushleft}
\hspace{60bp}$T_{1}^{1}=p_{r}=\frac{1}{2}[\frac{C(x,t)}{A(x,t)}(\frac{\partial\Phi}{\partial t})^{2}+\frac{1}{A(x,t)}(\frac{\partial\Phi}{\partial x})^{2}-2V(\Phi)]$
,\hspace{38bp}(25. c)
\par\end{flushleft}

\begin{flushleft}
\hspace{60bp}$T_{2}^{2}=p_{\bot}=\frac{1}{2}[\frac{C(x,t)}{A(x,t)}(\frac{\partial\Phi}{\partial t})^{2}-\frac{1}{A(x,t)}(\frac{\partial\Phi}{\partial x})^{2}-2V(\Phi)]$
.\hspace{36bp}(25. d)\vspace{1cm}
\par\end{flushleft}

\textbf{\large{}2.1. Boundary Conditions And Numerical Results}\textbf{
\vspace{0.5cm}
}

We can solve the (EKG) equations by considering the following Fourier
expansions {[}13{]}.

\begin{flushleft}
$\sqrt{k_{0}}\Phi(x,t)=2\sigma(x)cos(\omega t)$ ,\hspace{192bp}(26.
a)
\par\end{flushleft}

\begin{flushleft}
$A(x,t)=\stackrel[n=0]{n_{max}}{\sum}A_{n}(x)cos(n\omega t)=e^{\nu+\mu}=\hspace{90bp}e^{\nu_{0}+\mu_{0}}[I_{0}(\nu_{1}+\mu_{1})+2\stackrel[n=1]{n_{max}}{\sum}I_{n}(\nu_{1}+\mu_{1})cos(2n\omega t)]$
,\hspace{95bp}(26. b)
\par\end{flushleft}

\begin{flushleft}
$C(x,t)=\stackrel[n=0]{n_{max}}{\sum}C_{n}(x)cos(n\omega t)=e^{2\mu}=\hspace{150bp}e^{2\mu_{0}}[I_{0}(2\mu_{1})+2\stackrel[n=1]{n_{max}}{\sum}I_{n}(2\mu_{1})cos(2n\omega t)]$,
\hspace{136bp}(26. c)
\par\end{flushleft}

where $n_{max}$ is the Fourier mode at which the series are truncated.
It is remarkable that in Eq. (26 a) we have used the first term of
the scalar field , $\sqrt{k_{0}}\Phi(x,t)=\stackrel[n=1]{n_{max}}{\sum}\sigma(x)cos(n\omega t)$
, for simplicity, because the first term of the approximation is sufficient
to yield the main properties of oscillaton{[}16{]}. Non-singularity
and asymptotically flatness are two main characteristics of an oscillaton
which determine the boundary conditions. Solutions of (EKG) equations
must be regular at $x=0$, this means that $\sigma^{'}(0)=0$ and
$A(x=0)=0$. The latter condition is equivalent to $\nu_{0}(0)+\mu_{0}(0)=0$
, $\nu_{1}(0)+\mu_{1}(0)=0$ and $A_{n>0}(x=0)=0$. The conditions
$\sigma(\infty)=0$ and $A(x=\infty,t)=1$ are imposed by asymptotically
flatness solutions as well as $A_{0}(\infty)=1$, $A_{n>0}(\infty)=0$,
these conditions are equivalent to $\nu_{0}(\infty)+\mu_{0}(\infty)=0$
, $\nu_{1}(\infty)=0$ and $\mu_{1}(\infty)=0$. But the condition
of $\nu_{0}(\infty)+\mu_{0}(\infty)=0$ yields $C_{n>0}(\infty)=0$
and $C_{0}(\infty)\neq1$ because this variable determines the fundamental
frequency $\varOmega=\sqrt{C_{0}(\infty)}=e^{-\nu_{0}(\infty)}$as
an \textit{output value} after solving the oscillaton equations. The
next step is to choose $\sigma(0)$ as the\textit{ central valu}e
and $C_{n\geq2}$ for fulfilling of boundary conditions, Thereby we
obtain a set of eigenvalues for each central value. It is quotable
that Eqs. (26. b) and (26. c) confine the metric coefficients to even
nodes only. \vspace{1cm}

\textbf{\large{}3. Geodesic equations of motion} \vspace{0.5cm}

Spherically symmetric metric allows us to write the geodesics equations
for the metric (1). By using the geodesics equations which are in
the form of

\smallskip{}

\hspace{77bp}$\frac{d^{2}x^{\kappa}}{d\tau^{2}}=-\Gamma_{\mu\nu}^{\kappa}\frac{dx^{\mu}}{d\tau}\frac{dx^{\nu}}{d\tau}$
,\hspace{140bp}(27)

\smallskip{}

where $\Gamma_{\mu\nu}^{k}=\frac{1}{2}g^{k\eta}(-g_{\mu\nu,\eta}+g_{\nu\eta,\mu}+g_{\eta\mu,\nu})$
is affine connection, $\kappa$, $\eta$, $\mu$, $\nu$=$0$, $1$,
$2$, $3$ and $x^{\kappa}$, $x^{\mu}$ , $x^{\nu}$=$(t$ , $r$
, $\theta$ , $\phi)$, for a special case $(\theta=\frac{\pi}{2})$
we have:

\begin{flushleft}
\hspace{48bp}$\overset{..}{t}=-\frac{1}{2B}\frac{\partial B}{\partial t}\overset{.}{t}^{2}-\frac{1}{B}\frac{\partial B}{\partial r}\overset{.}{r}\overset{.}{t}-\frac{1}{2B}\frac{\partial A}{\partial t}\overset{.}{r}^{2}$,\hspace{115bp}(28.
a)
\par\end{flushleft}

\hspace{30bp}$\overset{..}{r}=-\frac{1}{2A}\frac{\partial B}{\partial r}\overset{.}{t}^{2}-\frac{1}{A}\frac{\partial A}{\partial t}\overset{.}{r}\overset{.}{t}-\frac{1}{2A}\frac{\partial A}{\partial r}\overset{.}{r}^{2}+\frac{r}{A}\overset{.}{\phi}^{2}$
,\hspace{90bp}(28. b)

\medskip{}

\hspace{30bp}$\overset{..}{\phi}=-\frac{2}{r}\overset{.}{r}\overset{.}{\phi}$
,\hspace{216bp}(28. c )

\medskip{}

where dot is derivative with respect to proper time($\tau)$. 

These equations are soluble numerically because it is not possible
to write the radial equation (28 b) as usual as $\frac{1}{2}\overset{.}{r}^{2}+U(r)=E$
where $\frac{1}{2}\overset{.}{r}^{2}$ is the kinetic energy, $U(r)$
an effective potential and $E$ the particles total energy. The condition
$L=0$ equivalent to $\phi=\phi_{0}=0$ means that a particle undergoes
a direct straight line free fall from its initial position ($r_{0},\phi_{0})$
. For $L\neq0$ a particle will undergo a variety of paths, on the
other hand for $L>0$ with increasing $\phi$, particles undergo in
the counter clock wise direction but for $L<0$ and with increasing
$\phi$ , particles undergo in the clock wise direction. Needless
to say that the motion depends on the central value of the radial
part of the scalar field , $\sigma(x=0)$ , the angular momentum $L$,
the initial radial position $r_{0}$ and initial radial speed $\overset{.}{r}_{0}$.
In turning point where $\overset{.}{r}_{0}=0$ , one of these two
options happen: either a particle goes away to infinity or oscillates
with a certain period and amplitude which could be found by solving
the geodesic equations numerically. Fig. 2. shows a path at which
a particle goes away to infinity gradually under the initial conditions
$r_{0}(\tau=0)=1.63$ , $\overset{.}{r}_{0}(\tau=0)=0$ and two different
values for momentum ,$L=1.5$ and $2.8$ for the case $\sigma(x=0)=0.3$
(at which singularity happens at point $x=1.62380$) for a range of
proper time $[\tau=0,40]$. As we can see with increasing $\tau$
, the path of the particles approaches to a closed circles. Changing
the radial position to $r_{0}(\tau=0)=2.63$ and momentum $L=1.5$
and $L=2.8$ but this time the proper time $\tau=100$ causes particles
undergo closed orbital. See Fig. 3. 

\bigskip{}

\includegraphics[width=5cm,height=5cm]{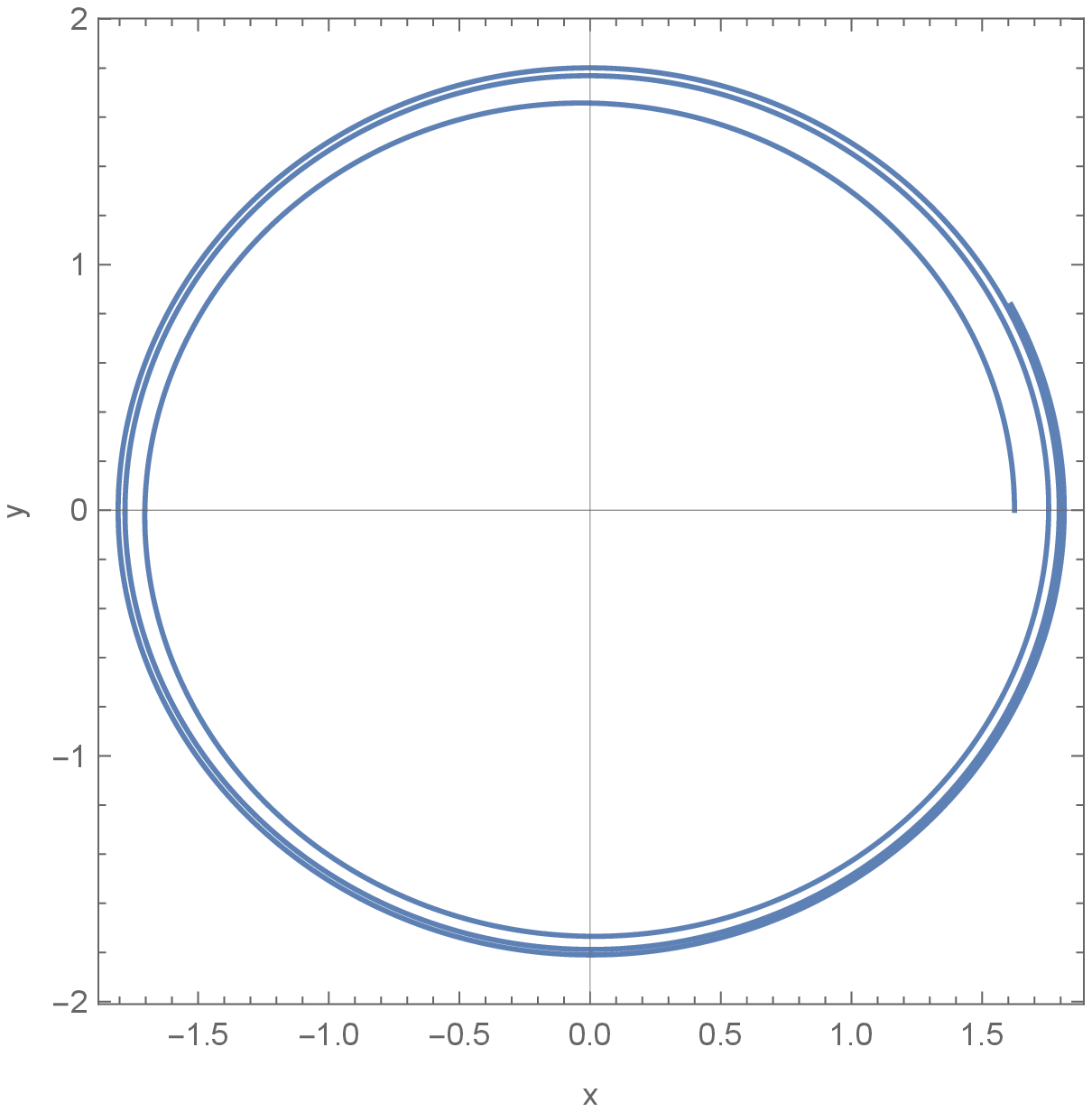}\includegraphics[width=5cm,height=5cm]{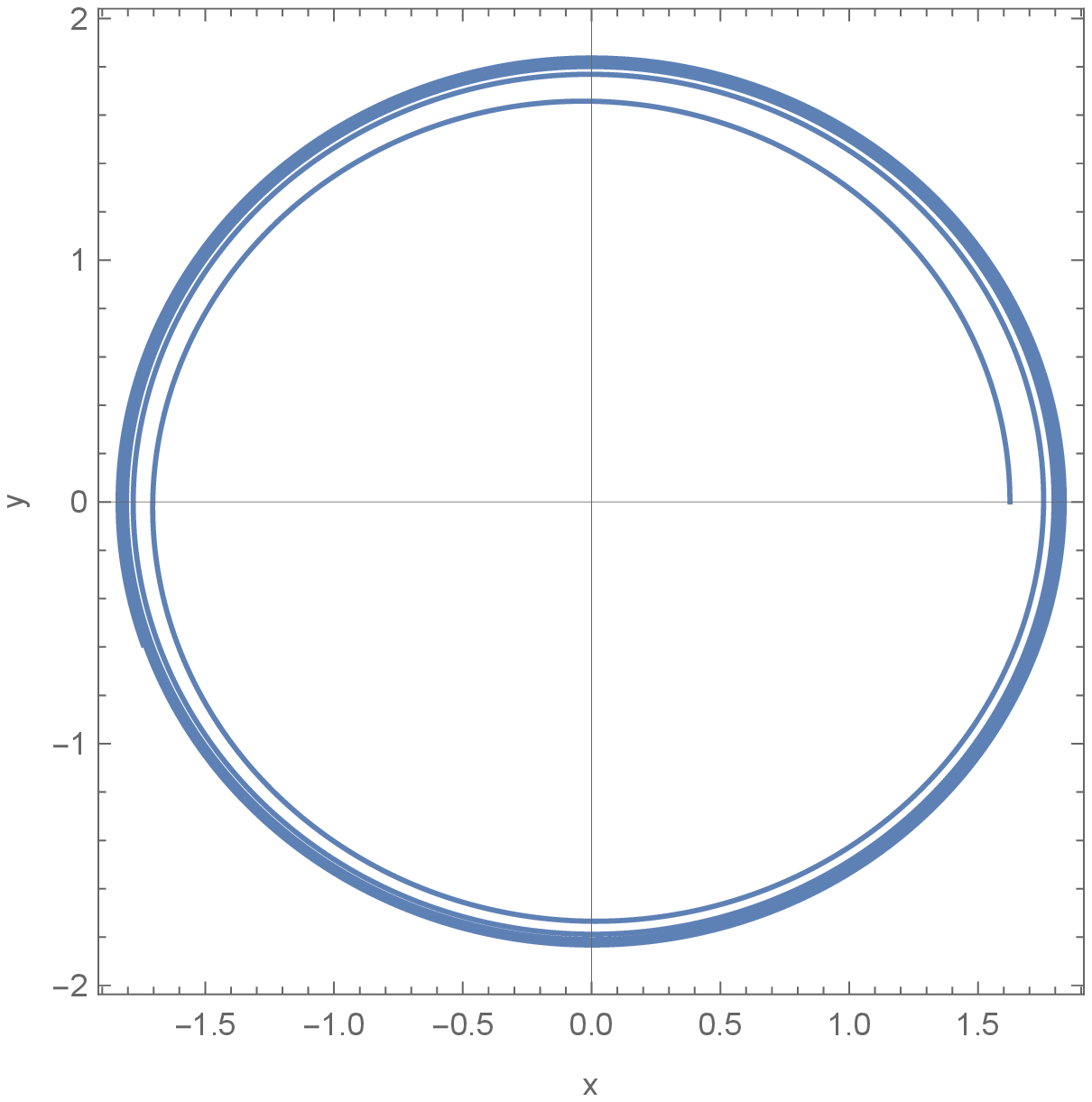}

Figs. 2. The orbits for particles under the boundary conditions: $\sigma(x=0)=0.3$,
$r_{0}(\tau=0)=1.62381$ , $\stackrel{.}{r}_{0}(\tau=0)=0$ and $L=1.5$
for left plot and $L=2.8$ for right plot at which proper time varies
from zero to 40.

\bigskip{}

\includegraphics[width=5cm,height=5cm]{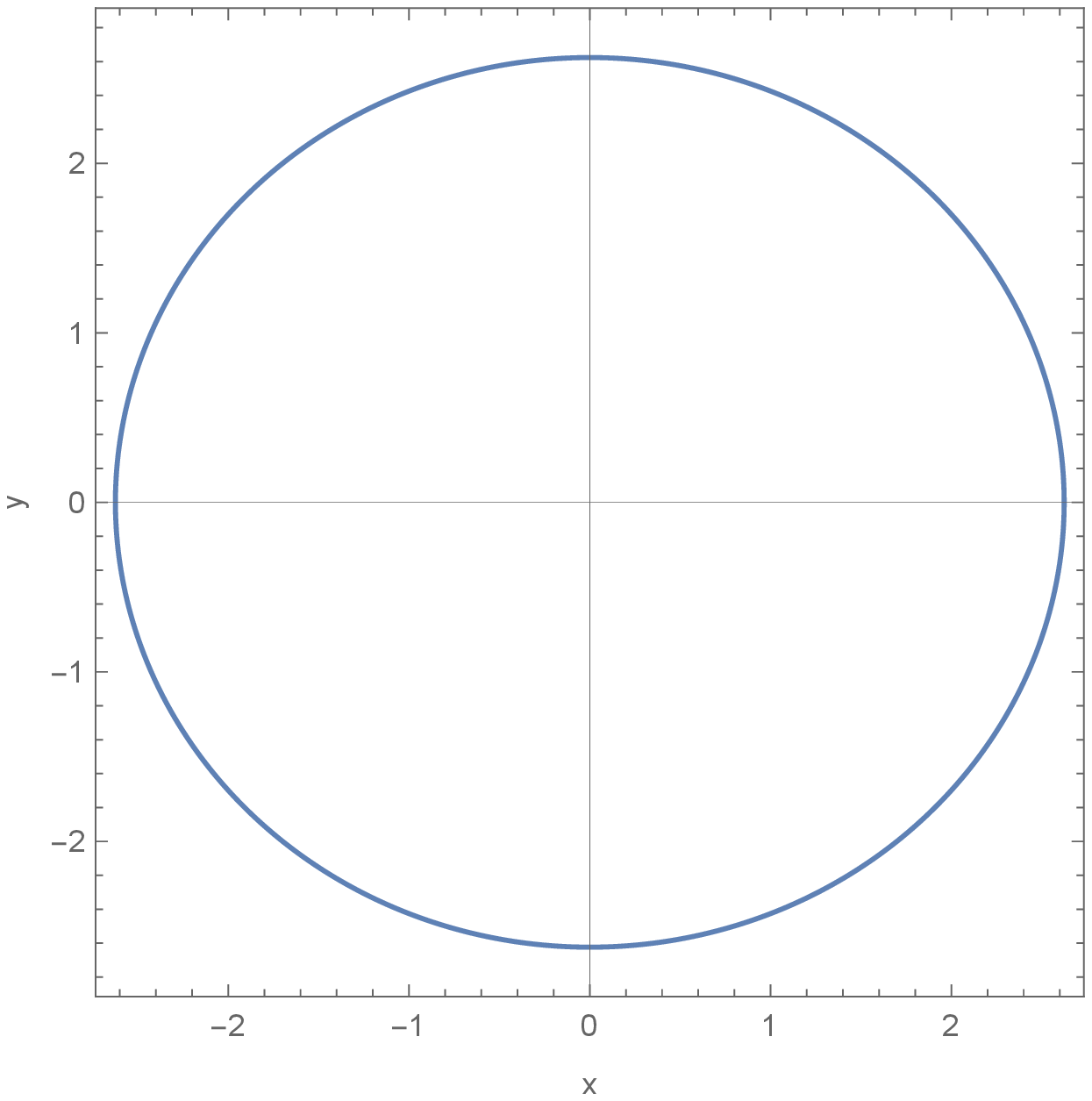}\includegraphics[width=5cm,height=5cm]{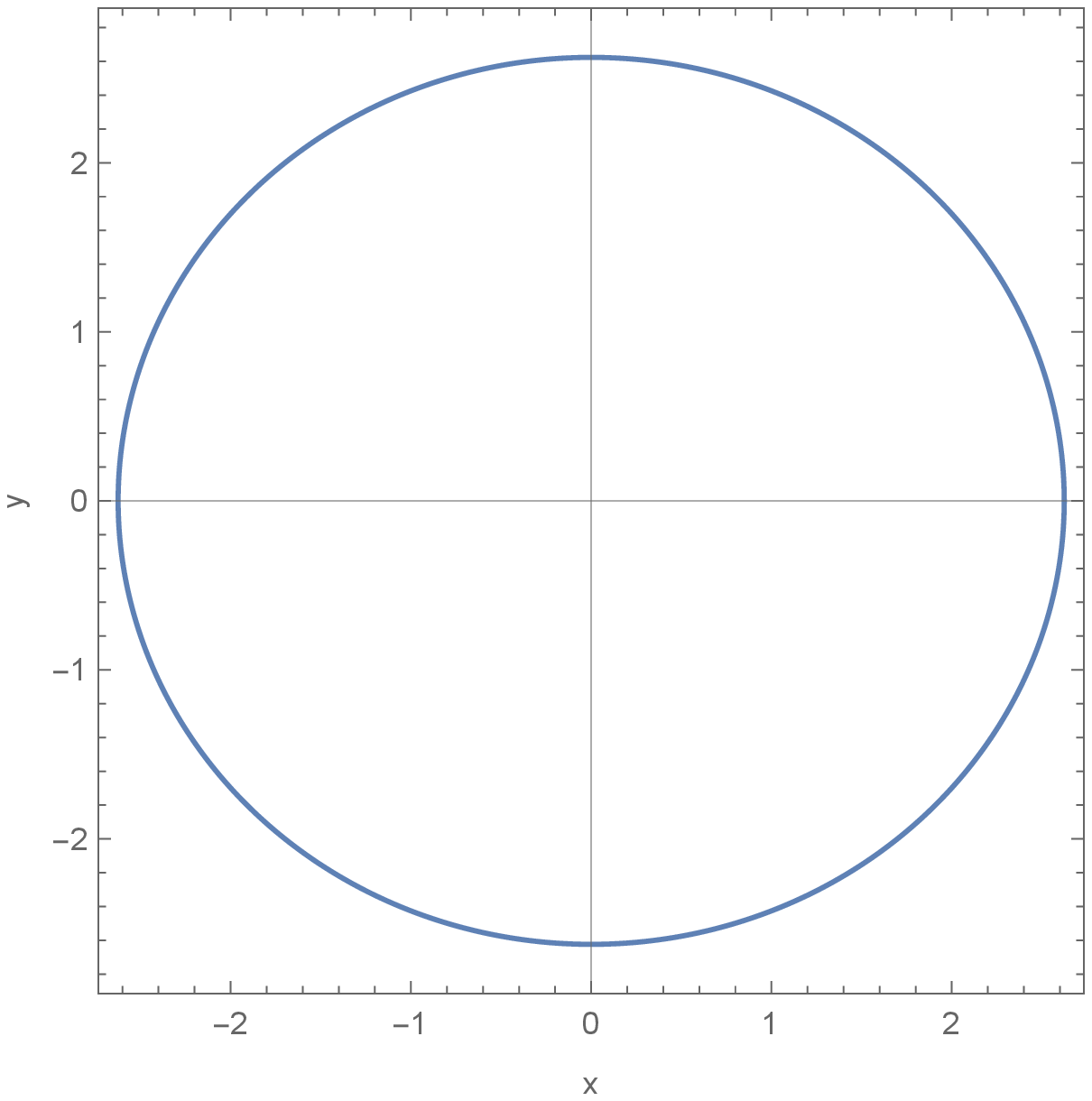}

Fig. 3. The orbits for particles under the boundary conditions: $\sigma(x=0)=0.3$,
$r_{0}(\tau=0)=2.62381$ , $\stackrel{.}{r}_{0}(\tau=0)=0$ and $L=1.5$
for left plot and $L=2.8$ for right plot at which proper time varies
from zero to 100.

\bigskip{}

Reiteration of the results mentioned above but this time for $\sigma(x=0)=0.7$
and the same $\lambda=-0.1$ show that singularity happens at point
$x=1.43696$ for metric functions, therefore for initial values of
$r_{0}(\tau=0)=1.44$ , $\overset{.}{r}_{0}(\tau=0)=0$ and two different
values for momentum , $L=1.5$ and $L=2.8$ ,when proper time varies
from zero to 40, the geodesics for real particles are shown in Fig.
4 and Fig. 5.

\bigskip{}

.\includegraphics[width=5cm,height=5cm]{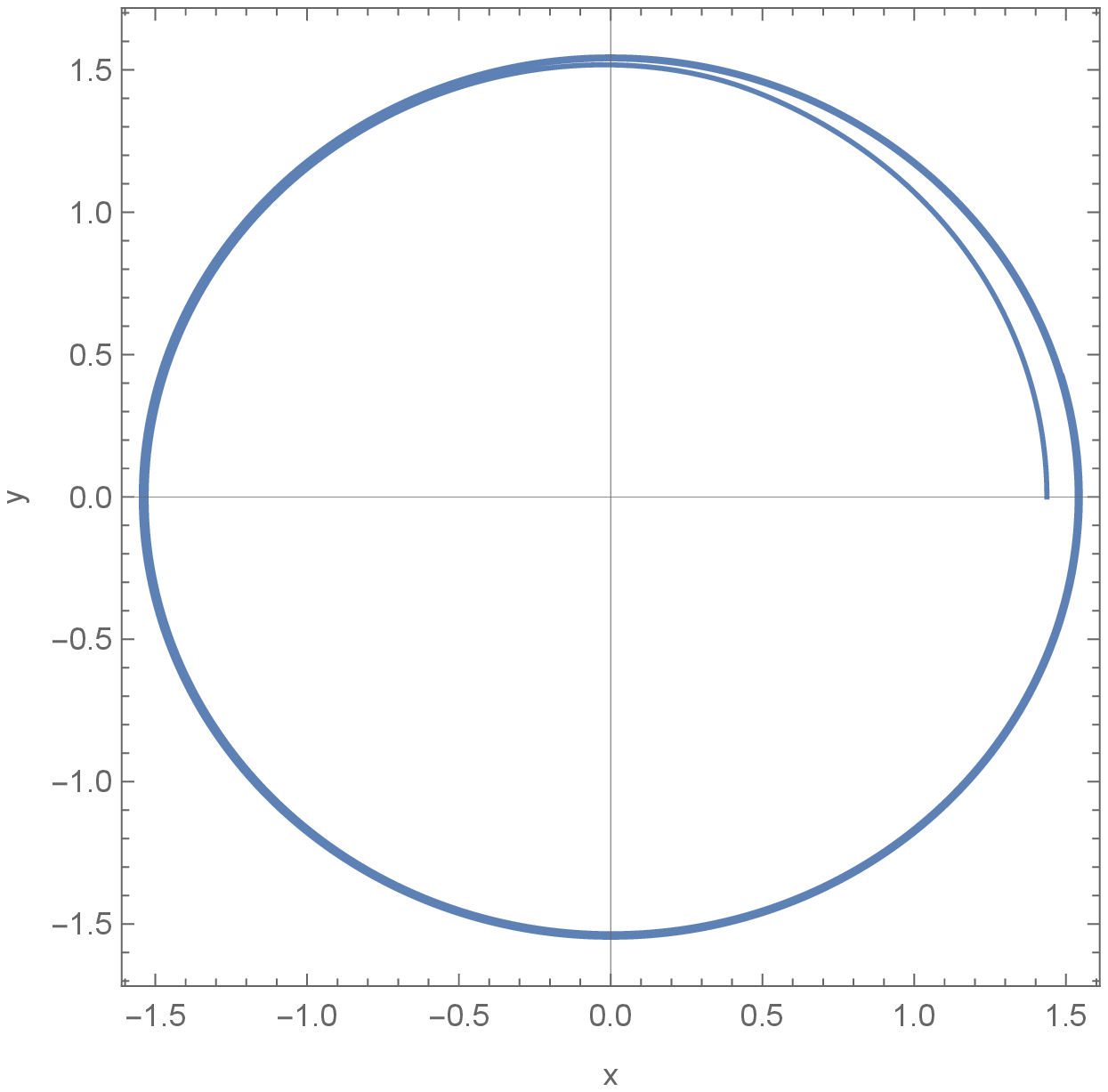}\includegraphics[width=5cm,height=5cm]{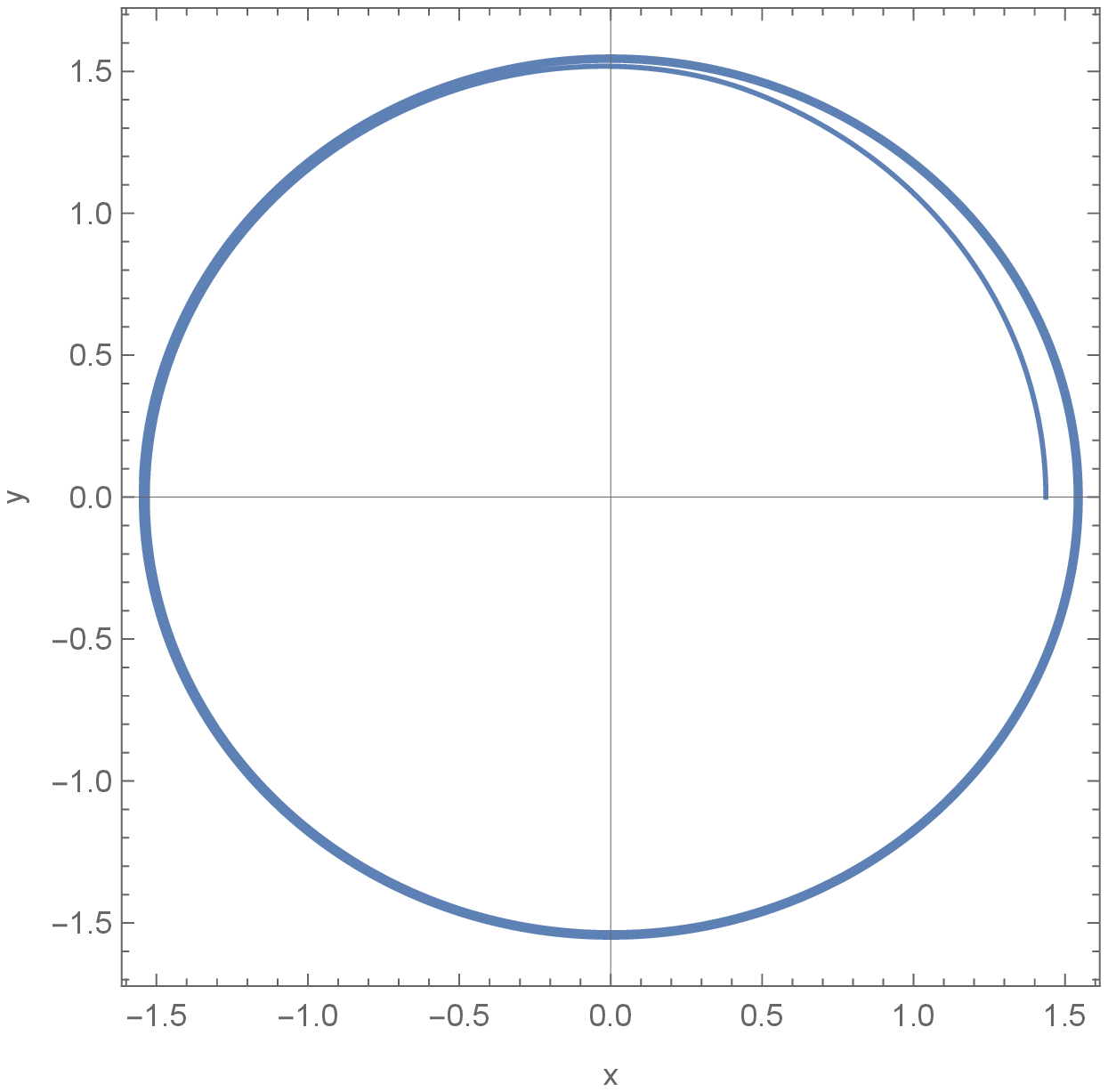}

Fig. 4. The orbits for particles under the boundary conditions: $\sigma(x=0)=0.7$,
$r_{0}(\tau=0)=1.43697$ , $\stackrel{.}{r}_{0}(\tau=0)=0$ and $L=1.5$
for left plot and $L=2.8$ for right plot at which proper time varies
from zero to 40.

\bigskip{}

\includegraphics[width=5cm,height=5cm]{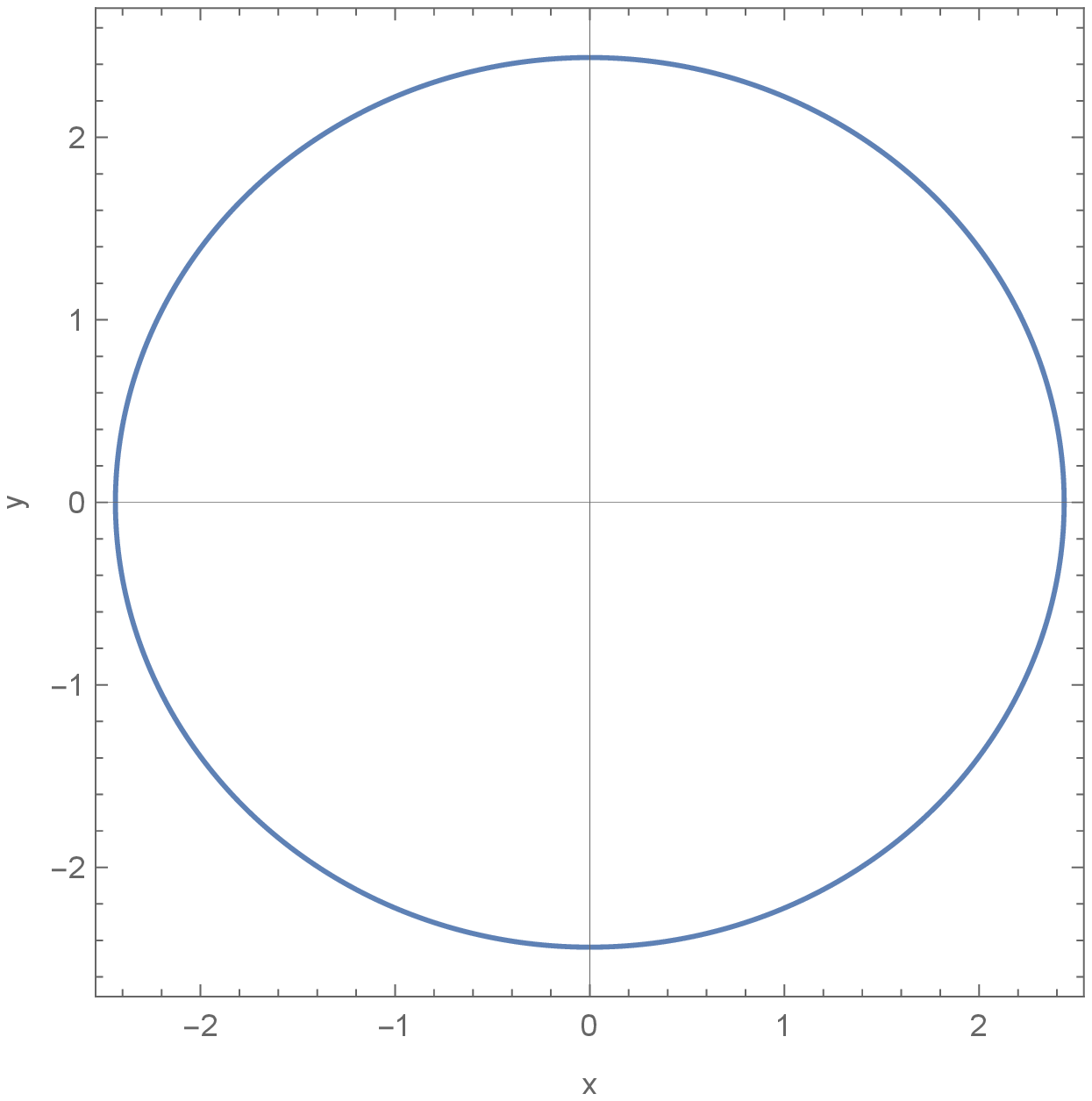}\includegraphics[width=5cm,height=5cm]{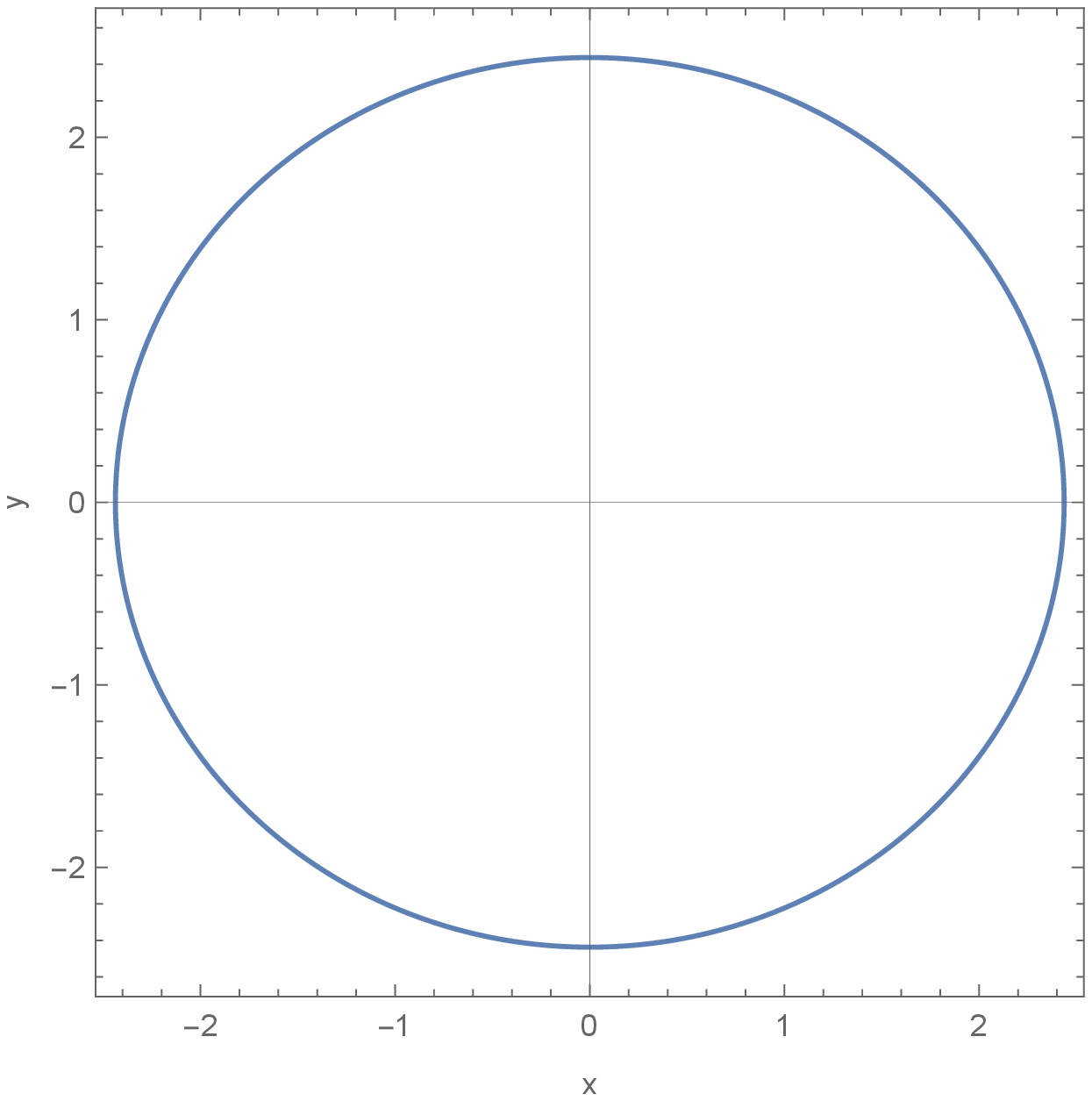}

Fig. 5. The orbits for particles under the boundary conditions: $\sigma(x=0)=0.7$,
$r_{0}(\tau=0)=2.43697$ , $\stackrel{.}{r}_{0}(\tau=0)=0$ and $L=1.5$
for left plot and $L=2.8$ for right plot at which proper time varies
from zero to 40.

\vspace{1cm}

\textbf{\large{}4. conclusions}{\large \par}

\vspace{0.5cm}

In this study we presented the simplest approximation for solving
a minimal coupled Einstein-Klein-Gordon equations for a spherically
symmetric oscillating soliton object called oscillaton endowed with
an exponential scalar field potential $V(\Phi)=V_{0}e^{\lambda k_{0}\Phi}$
and an harmonic time-dependent scalar field $\Phi(r,t)=2\sigma(r)cos(\omega t)$
. By taking into account the Fourier expansions of differential equations
and with regard to the boundary conditions which require the non singularity
and asymptotically flatness, the solutions ( the metric functions)
are obtained numerically. It is more convenient to fit these solutions
which are in the form of Interpolation Functions to new functions,
In present study this work has been done to order four, although better
results are obtained by taking into account the higher order of fitting.
By determination of the metric functions, the metric coefficients
are also obtained easily. As we know the metric coefficients are the
main tools for dealing with the geodesic equations according to $\frac{d^{2}x^{k}}{d\tau^{2}}+\Gamma_{\mu\nu}^{k}\frac{dx_{\mu}}{d\tau}\frac{dx_{\nu}}{d\tau}=0$
. Solving these equations based on initial boundary conditions and
plot the answers in the polar coordinates show the trajectory of the
real particles . This work has been done at points very close to what
is called the singularity for one of the metric functions for two
different central values of $\sigma(x=0)=0.3$ and $0.7$ and momentum
values $L=1.5$ and $2.8$ at turning points where $\overset{.}{r_{0}}=0$
, the particle$^{,}$s speed, is equal to zero. Results show that
for radial points outside the oscillaton and close to the singularity
point the particles$^{,}$ trajectory gradually increase or decrease
for positive and negative values of the momentum respectively and
for points sufficiently far from the oscillton the particles$^{,}$path
are closed orbits. Despite the presented results in this study, there
are some problems that should be investigated elsewhere in the future
works. 1. Under which conditions and for which values of $\sigma$
, $L$ , and other initial values the radial coordinates of particles
oscillate in time. 2. If the radial coordinates of the particles oscillate
in time then the amplitude of oscillation and the period of oscillations
are desirable.

\end{document}